\documentclass[12pt,a4paper]{article}
\pdfoutput=1
\newcommand{\be}{\begin{equation}}
\newcommand{\ee}{\end{equation}}
\newcommand{\bea}{\begin{eqnarray}}
\newcommand{\eea}{\end{eqnarray}}
\newcommand{\beas}{\begin{eqnarray*}}
\newcommand{\eeas}{\end{eqnarray*}}
\newcommand{\ba}{\begin{array}}
\newcommand{\ea}{\end{array}}
\newcommand{\nn}{\nonumber}

\newcommand{\bt}{\begin{table}}
\newcommand{\ve}{\varepsilon}
\newcommand{\vsi}{\varsigma}

\newcommand{\al}{\alpha}
\newcommand{\ga}{\gamma}
\newcommand{\bmu}{\bar\mu}
\newcommand{\bnu}{\bar\nu}
\newcommand{\bla}{\bar\lambda}
\newcommand{\brho}{\bar\rho}
\newcommand{\bm}{m}
\newcommand{\bn}{n}
\newcommand{\bl}{l}
\newcommand{\br}{r}
\newcommand{\bk}{k}
\newcommand{\bs}{s}
\renewcommand{\bnu}{n}
\renewcommand{\bmu}{m}
\renewcommand{\bla}{l}
\renewcommand{\brho}{r}
\newcommand{\Ga}{\Gamma}	
\newcommand{\de}{\delta}
\newcommand{\De}{\Delta}
\newcommand{\ka}{\kappa}
\newcommand{\la}{\lambda}
\newcommand{\La}{\Lambda}
\newcommand{\na}{\nabla}

\newcommand{\si}{\sigma}

\begin{document}

\title{\bf 
Emergent gravity, violated relativity  and dark matter}
\author{Yury~F.~Pirogov\footnote{E-mail: yury.pirogov@ihep.ru }
\\
\small{\em Theory Division, Institute for High Energy Physics,  Protvino, 
Moscow Region, Russia}
}
\date{}
\maketitle

\begin{abstract}
\noindent  
The nonlinear affine Goldstone model  of  the emergent gravity, built on  the
nonlinearly realized/hidden affine symmetry, is  concisely 
revisited. Beyond General Relativity, 
the explicit violation of   general invariance/relativity,
under preserving general covariance,  is exposed. 
Dependent on  a nondynamical affine connection,
a generally covariant second-order effective Lagrangian for metric gravity
is worked out, with the  general relativity violation   and  the
gravitational dark matter serving as the signatures of emergence.



\end{abstract}

\section{Introduction}
It is widely accepted    nowadays that General Relativity (GR) may be  just 
(a piece of)  an effective field theory of gravity  to be  ultimately superseded
at the
high energies by a more fundamental/underlying  theory.  At that,  the
conventional  metric  gravity could cease to be a priori existent,   but, 
instead, would become   an emergent/induced phenomenon. 
A~lot of the drastically different approaches towards the emergence of 
gravity and space-time 
is presently conceivable.\footnote{For recent  surveys  of the emergent gravity
and space-time, see, e.g.~\cite{Sind,Carl}.} In this paper, we  work
out an  approach to the goal  treating the   gravity as  an
affine Goldstone phenomenon in the framework of the effective field theory.

As a herald  of  an unknown  high-energy theory there typically   serves at
the lower energies  a nonlinear model.   Being  based
on a nonlinearly realized/hidden symmetry, remaining linear on  an
unbroken subgroup, such a model  could encounter  in a concise  manner for the
spontaneously/dynamically broken symmetries of the
fundamental theory. Inevitably, this occurs at the cost of  more
uncertainty and  a  partial  loss  of content.  For the 
global continuous internal symmetries,  the nonlinear model framework was
developed
in~\cite{NM1,NM2}. This approach proved to be extremely  useful for
studying, e.g.,  the so-called chiral model and played an important role in
the advent of QCD as the true fundamental theory of strong interactions. 

One might thus  naturally expect that in the quest for an underlying theory of
gravity GR should first be substituted by  a nonlinear model.
As such a model for gravity,  aimed  principally at  
reconstructing GR, there was originally proposed  the model   based
on the nonlinearly realized/hidden  affine symmetry,  remaining
linear on the unbroken Poincare subgroup~\cite{HAS1,HAS2}.\footnote{For a
fiber bundle formalism, cf.\  \cite{Sard}.} 
In the context of emergence of the gravity and space-time, the model was 
elaborated in~\cite{Pir1}. 
At that, reproducing  GR the model may well include the 
general invariance/relativity
violation~\cite{Pir1}--\cite{Pir4}.\footnote{The term  general covariance
violation used in \cite{Pir1}--\cite{Pir3a}  is  to be more  appropriately
substituted  by  the general invariance/relativity violation~\cite{Pir2,Pir4}.}
To this end, one  should  envisage in a field theory  two kinds of  fields
--  the dynamical/relative 
and nondynamical/absolute ones --  and, respectively,  two kinds 
of the diffeomorphism symmetries. First, the
kinematical  symmetry -- the covariance --  which restricts    the
mathematical form of  the  theory. 
Second,  the dynamical symmetry -- the 
invariance/relativity -- which serves as a gauge symmetry
for  gravity  determining   the physical content of the latter.
In GR, without nondynamical fields, these notions  coincide, both being the
general ones. But beyond GR, in the  presence of nondynamical  fields,  
the notions differ~\cite{Pir2,Pir4}.\footnote{For a discussion of the general
covariance vs.\ general invariance, cf.\ also
\cite{And}.}
For consistency, the  general covariance should better
be preserved. On the other hand,   the GR violation  may well  take place,
serving as a source of the gravitational  dark matter~(DM).  In a simplest
case, such
a scenario was worked out for  the well-defined theory of gravity minimally
violating GR to the  unimodular relativity,  with the
scalar-graviton DM~\cite{Pir1a}--\cite{Pir4}.
Extending  this scenario  to  other types of  the GR violation and  
gravitational DM would  thus be  urgent.

In this paper,   the model of emergent gravity based on  the nonlinearly
realized/hidden affine symmetry
 --  the nonlinear affine Godstone  model -- 
is systematically revisited. 
To allow for the GR violation, two kinds of coordinates --  the
absolute/background and  relative/observer's  ones --  are envisaged.
A generally covariant second-order  effective
Lagrangian for metric gravity,
dependent on a nondynamical affine
connection,  is consistently worked out  in the 
 most general fashion, with a limited version  discussed  in more detail. 
The model
is proposed as a prototype for the emergent gravity and space-time,
with the GR violation and the gravitational DM serving  as the
signatures of emergence.

\section{Nonlinear realizations and emergent gravity}

\subsection{Spontaneous symmetry breaking and nonlinear realizations}

To begin with,  let us shortly recapitulate the techniques  of the nonlinearly
realized/hidden symmetries. Let a  global continuous internal symmetry $G$, with
the dimension $d_G$, be
spontaneously/dynamically   broken,  $G\to H$,  to some $d_H$-dimensional 
subgroup $H\subset G$.  Let  $K=G/H\subset G$ be the respective
$d_K$-dimensional  (for definiteness, left) 
coset space consisting of  the (left) coset elements $k\in K$. Then  any group
element $g\in G$ admits a unique (at least in a vicinity of
unity)  decomposition  $g=k
h$, with   $k\in K$ and    $h\in H$. 
Henceforth under the action of a group element $g_0$, one should  get $g_0 k= k'
 h'(g_0,k)$. The group  $G$ thus acts on  $k$ by means of 
the transformations $k\stackrel{g_0}{\to }k'=g_0 k
h'^{-1}(g_0,k)$ dependent, generally,  on $k$ (henceforth the term nonlinear).
Mapping a flat space $R^d$ onto
$K$, $R^d \to K$,   defines  on $R^d$ a coset-valued field $k( \xi)\in  
K$, $ \xi\in R^d$.  This induces a nonlinear realization of 
$G$ on $K$.
Restricted by the unbroken subgroup  $H$,  i.e., under  $g_0=h_0$,  the
nonlinear realization of $G$
is  to be a  usual linear representation of  $H$, 
$k\to k'=h_0 k h^{-1}_0$, with $h'(h_0,k)=h_0$, and thus   $h'(I,k)=I$.
Putting
$k=\exp (\sum_i \pi_i X_i)$,  $i=1,\dots, d_K$,  $ d_K=d_G - d_H$, with $X_i$
being the broken generators of $G$, one
can treat the $d_K$-component field $\pi$  as a  Goldstone
boson emerging under the global  symmetry breaking. 
Due to  the isomorphism $G  \simeq   K   \otimes   H$ (at least in a vicinity of
unity), one can substitute  a   coset  element $k$
by  its equivalence class $\ka$ obtained from $k$ through the (right)
multiplication by an  arbitrary $h\in H$. At that, 
the nonlinear realization gets linearized as  $\ka\stackrel{g_0}{\to
}\ka'=g_0 \ka h^{-1}$,  modulo  an  arbitrary  $h( \xi)\in
H_{\rm loc}$ independent of  $\ka $.  And v.v., fixing
a gauge for $H_{\rm loc}$ results in imposing the $d_H=d_G-d_K$ restrictions 
on $\ka$ and choosing thus a nonlinear realization
$k\stackrel{g_0}{\to}k'= g_0 k  h'^{-1}(g_0, k)$.  Hence, being equivalent to
the linear representation,  all the nonlinear realizations for breaking
$G\to H$ are equivalent among themselves in the effective field theory sense.
At that, the linearization  of a hidden symmetry  may be more advantageous   as
embodying on par all the  equivalent nonlinear realizations.

\subsection{Gravity as an affine Goldstone phenomenon}

In the case at hand, an underlying  theory of   gravity is to be originally 
invariant under the global affine group
$G=IGL(d, R)$, $d=4$.\footnote{The following consideration is formally 
independent of $d\ge 2$.} 
Eventually, the symmetry spontaneously/dynamically  
brakes  down to the Poincare  one, assumed to be exact:
\be
G=IGL(d,R)\to H= ISO(1,d-1).
\ee
A putative mechanism of such a  breaking  is  beyond the scope of  the model.
According to general theory,  the  breaking results in  the  nonlinear
realization of the  affine symmetry  on the coset space
$G/H = IGL(d,R)/ISO(1,d-1)$,  with the 
$d(d+1)/2$-component coset
elements. The Goldstone boson  of  the   respective nonlinear realization
is to  be  treated as a  primary   gravity field.
To consistently apply  the  nonlinear  realization technique  to 
such a global  external  symmetry,  a two-stage procedure  is to be
implemented, starting  from  a flat affine background and extending  then to  
a curved one.

\subsection{Flat affine background}

First, let ${\cal R}_d\simeq R^d$  be a $d$-dimensional homogeneous   space,  
with the affine group as the group of motions, ${\cal R}_d \to {\cal R}_d$.   By
default, ${\cal R}_d$ admits the globally affine  coordinates
$\xi^m\in R^d$, $m=0,1,\dots, n-1$,\footnote{Being here  just a notation, the
index $m=0$ is  understood  to 
subsequently compile  with the unbroken Lorentz subgroup.}     
undergoing  the  affine transformations:
\be
\xi^m \stackrel{(A, a)}{\to} \xi'^m= \xi^n A^{-1 m}_n+a^m,
\ee
with the  arbitrary constant  parameters $A_n{}^m$ and 
$a^m$ for the (reversible)  linear deformations and translations,
respectively.\footnote{Remaining unbroken, 
the translation part of the  symmetry  is omitted in what follows.}
This  space  will serve  as the 
representation one  for constructing the nonlinear model.
In accord with the general formalism  there are two  modes  for
realization of the hidden affine symmetry:  the
nonlinear and linearized  ones.

\subsubsection{Pseudo-symmetric nonlinear realization} 
A  coset element 
$\vartheta_m{}^a$, $a=0,1,\dots, d-1$, may
uniquely  be chosen  to be   pseudo-symmetric,~i.e.,
\be
 \eta^{am}\vartheta_m{}^b =  \eta^{bm}\vartheta_m{}^a
\ee
(at least in a  suitable neighbourhood of  $\vartheta_m{}^a=\de^a_m$, where
this condition  is evidently fulfilled).
Here,  $\eta^{ab}$ (and $\eta_{ab}$)  is the invariant under $SO(1,d-1)$
Minkowski symbol,  by which the globally Lorentzian indices $a, b$, etc., are
manipulated. Under $A\in GL(d,R)$ the coset element
should transform nonlinearly as
\be
\vartheta_m{}^a(\xi)\stackrel{A}{\to} \vartheta'_m{}^a(\xi')=A_m{}^n
\vartheta_n{}^b(\xi)\La'^{-1}_b{}^a(A,\vartheta),
\ee
with $\La' \in SO(1,d-1)\subset GL(d,R)$
chosen so to retain the  pseudo-symmetry after 
action of  $A$. At that, due to $\La_{ab}=\La^{-1}_{ba}$ there automatically 
fulfills the linearity condition $ \La'(\La,\vartheta)=\La$ for any $\La\in
SO(1,d-1)$.\footnote{Moreover, the dilatations  $R$ are represented linearly,
too, with the trivial  compensating factor, $\La'(R,\vartheta)=I$.}
Present  the symmetric Lorentz tensor $\vartheta^{ab}\equiv
\eta^{am}\vartheta_m{}^b$ in terms of a symmetric
tensor  $h^{ab}$ as $\vartheta\equiv \exp (h/2)$, where, e.g.,
 $(h h)^{ab}= h^{ac} h^{bd}\eta_{cd}$, etc. With $H=SO(1,d-1)$ serving as a
classification group, one can treat $h^{ab}$, with   
$d(d+1)/2$ independent components,   as a tensor Goldstone boson emerging under 
the breaking of the global affine symmetry to the Poincare  one.

\subsubsection{Locally Lorentzian linear  representation} 
The affine Goldstone model  gets simplified with   the
nonlinear realization being  linearized in terms of a 
$d^2$-component frame-like field $\vartheta_m^\al$,
$\al=0,1,\dots, n-1$  (and its  inverse  $\vartheta^m_\al$). The
field  transforms  under $A\in GL(d,R)$~as 
\be
\vartheta_m^\al(\xi)
\stackrel{A}{\to} \vartheta'^\al_m(\xi')=A_m{}^n
\vartheta_n^\beta(\xi)\La^{-1}_\beta{}^\al(\xi)
\ee
(and likewise for  $\vartheta_m^\al$), modulo  an  arbitrary
$\La_\al{}^\beta(\xi) \in SO(1,d-1)_{\rm loc}$  
satisfying $ \La_{\al\beta}=\La^{-1}_{\beta\al}$, with the 
invariant $\eta_{\al\beta}$.
Due to invariance under  $SO(1,d-1)_{loc}$, with the $d(d-1)/2$ local
parameters, the number of the independent  components remains in fact 
the same $d(d+1)/2$.
Explicitly, one can  impose a  Lorentz gauge by projecting
$\vartheta_m^\al\to \vartheta_m{}^a = \vartheta_m^\beta
\La^{-1}_\beta{}^a(\vartheta)$  so that
$\vartheta_m {}^a$  becomes pseudo-symmetric. Thus, the two realization  modes,
the nonlinear and linearized ones,  are  
equivalent. However, the  linear  representation may be more 
advantageous due to grasping all the equivalent nonlinear realizations.

Restricting ourselves to the pure gravity, introduce the  
generic action  element produced by an  infinitesimal neighbourhood
$d^d\xi$ of  a reference point $\Xi$ as follows:
\be\label{mismatch}
d S    =  {\cal L}_g \, d^d\xi \equiv    L_g (\vartheta_m^\al,  \partial_n 
\vartheta_m^\al ,\dots) |\mbox{\rm
det} (\vartheta_m^\al)|d^d \xi,
\ee 
with $d S$ being invariant relative to $GL(d,R)\otimes
SO(1,n-1)_{\rm loc}$.
For the pure gravity, the field $\vartheta_m^\al$   
may be converted into  the  symmetric Lorentz-invariant second-rank 
affine tensor $g_{mn}$ with $d(d+1)/2$ independent degrees of freedom:
\be
g_{mn}(\xi)=\vartheta_m^\al \eta_{\al\beta}
\vartheta_n^\beta.
\ee
In particular, one has $ |\mbox{\rm
det} (\vartheta_m^\al)| =\sqrt{ |{\rm det}( g_{mn})|}$. 
This tensor  will    eventually be
treated as the emergent  metric being  absent prior to the affine
symmetry breaking.  
Transforming the result to the arbitrary curvilinear coordinates  $x^\mu$ on
${\cal R}_d$  and
integrating over  ${\cal R}_d$ one would get the nonlinear model of
gravity on the  flat affine background.
To account for a more general  topology, one  should go over to a
curved affine background, with  the action
element (\ref{mismatch}) used as a flat counterpart.

\subsection{Curved affine  background}

Let now ${\cal M}_d$  be a  $d$-dimensional differentiable ``world'' 
manifold marked   with some  ``observer's''
coordinates $x^\mu\in R^d$, $\mu=0, 1,\dots, n-1$.\footnote{The
index $\mu=0$ here is  just a notation acquiring the physical  meaning
after the emergence of metric.} 
Let  ${\cal M}_d$ be moreover  endowed with a nondynamical 
affine connection $\hat\Ga_{\mu\nu}^\la(x)$,   
free of torsion, $\hat\Ga_{\mu\nu}^\la-\hat\Ga_{\nu\mu}^\la=0 $.
Such a  background affine structure is to be formed by an underlying
theory on par with  the affine symmetry breaking.\footnote{This reflects the
assumption of the absence of a    prior metric  structure on an underlying
level.  Only an affine texture of  the background is supposed.} 
In a vicinity of a fixed, but otherwise arbitrary point $X$
the connection may be decomposed  as follows 
\be\label{decomp}
\hat\Ga_{\mu\nu}^{\la}( x) =  \hat\Ga_{\mu\nu}^{\la}( X)  +   \frac{1}{2}
\hat R_{\mu
\rho\nu}^{\la}(X)(x-X)^{\rho}+ {\cal  O}((x-X)^2),
\ee
where  $\hat R_{\mu \rho\nu}^{\la}(X)$ is the background  curvature
tensor in the  reference point $X$. 
In a patch  around 
$X$, choose on    ${\cal M}_d$  some  coordinates
$ \xi_X ^{m}= \xi_X^{m} (x)$ (having the inverse  $x^{\mu}= 
x^{\mu} (\xi_X )$)  so that conventionally 
\be\label{trans}
\hat\Ga_{mn}^{l}(\xi_X )={e} ^\mu _{m} 
{e}^\nu _{n} {e} ^{l}_{\la}(x) \Big(  \hat\Ga_{\mu\nu}^\la(x) - 
{e}^{\la}_{r} (x) \partial^2 \xi_X^{r}/\partial x^\mu\partial x^\nu
\Big),
\ee
where ${e} ^{m}_{\mu}(x) =  \partial \xi_X ^{m}/\partial
x^{\mu}$ and   ${e} _{m}^{\mu}(x) =  \partial x^{\mu}/\partial 
\xi_X^{m}|_{\xi_X =\xi_X (x)}$.  
Differentiating the reversibility relations $\xi_X ^{m}(x(\xi_X ))= \xi_X
^{m}$ and 
$x^{\mu} (\xi_X (x))=x^{\mu}$
one gets  ${e} ^{\bmu}_\la {e} _{\bnu}^\la =  \de ^{\bmu}_{\bnu}$
and
$ {e}_{\bla}^\mu {e} ^{\bla}_\nu =  \de_\nu^\mu$. 
More particularly, adjust  the coordinates $\xi ^{m}_X$   as
follows:
\be
\xi^{\bla}_X = \Xi_X^{\bla} +{e} ^{\bla}_\la (X)
\Big((x-X)^\la+\frac{1}{2} \hat\Ga_{\mu\nu}^{\la}(X)  (x-X)^\mu 
(x-X)^\nu\Big) +{\cal  O}((x-X)^3),
\ee 
implying
\be\label{xbarx}
{e} _{\bla}^\la (X)  \partial^2 
\xi^{\bla}_X/\partial x^\mu\partial x^\nu |_{x=X}
= \hat
\Ga_{\mu\nu}^\la(X) .  
\ee
In view of (\ref{trans}) one thus gets $ \hat\Ga_{\bmu\bnu}^{\bla}(
\Xi_X)=0$. Moreover,  in view  of (\ref{decomp}) for coordinates $\xi^m_X$
one has 
\be\label{locaff}
\hat\Ga_{\bmu\bnu}^{\bla}(\xi_X ) =\frac{1}{2} \hat R_{\bmu
\brho\bnu}^{\bla}(\Xi_X)(\xi_X-\Xi_X) ^{\brho}+ {\cal 
O}((\xi_X-\Xi_X) ^2).
\ee
The  manifold ${\cal M}_d$  looking   in the coordinates $\xi _X^{\bmu}$
approximately flat around $\Xi_X$,  call such coordinates  the locally
affine ones  in the point  $X$. In these coordinates,  project a patch of ${\cal
M}_d$ around $X$ onto  the representation space
${\cal R}_d$ (associated with the tangent space in $X$)  through $\xi^m
-\Xi^{\bmu}=
\xi_X^m -\Xi_X^m +{\cal O}((\xi_X- \Xi_X)^2)$.
This maps in the leading approximation the action element (\ref{mismatch})
from ${\cal R}_d$ onto ${\cal M}_d$.
Transform then   the local result around  $\Xi_X$ from coordinates $\xi_X^m$
to the arbitrary observer's  coordinates $x^\mu$
by means of   substitutions $\vartheta_{\bm}^\al={e}
_{\bm}^\mu\vartheta_\mu^\al$,
$\partial_{m}= {e} ^\mu_{ m}\partial_\mu$  and $d^d
\xi_X =|{\rm det}({e} _\mu^{m})| d^d x$,  with  frames
$\epsilon^\mu_m(x)$ and $\epsilon_\mu^m(x)$.
Integrating finally  over ${\cal M}_d$  one gets
the generic gravity action (redefining $X\to x$) as follows:
\be\label{biframe}
S=  \int L_g (\vartheta_\mu ^\al,  \partial_\nu\vartheta_\mu ^\al,  \dots; 
\hat\Ga_{\mu\nu}^\la )
|\mbox{\rm det} (\vartheta_\mu^\al)|d^d x. 
\ee
Evidently, the   locally Lorentzian
frame  $\vartheta^\al_\mu$   satisfies    the
reversibility relations $\vartheta^\al_\mu 
\vartheta_\beta^\mu =  \de ^\al_\beta$ and $\vartheta_\al^\mu \vartheta^\al_\nu
= \de_\nu^\mu$. 
Due to  $\vartheta^\al_\mu= {e} ^{\bmu}_\mu
\vartheta_{\bmu}^\al $   the frame transforms under a diffeomorphism $x^\mu \to
x'^\mu=x'^\mu(x)$ as 
\be
\vartheta^\al_\mu(x)\to
\vartheta'^\al_\mu(x')=\frac{\partial
x^\nu}{\partial x'^\mu}  
\vartheta^\beta_\nu(x)\La^{-1}_\beta{}^\al(x),
\ee
modulo an arbitrary 
$\La  \in SO(1,d-1)_{\rm loc}$ (and likewise for $\vartheta_\al^\mu=  {e}
_{\bmu} ^\mu \vartheta^{\bmu}_\al $). 
The action~(\ref{biframe}), as originated from  (\ref{mismatch}),  is
to be invariant relative to the local
Lorentz  transformations, being at the same time generally covariant. 
At that, the Lorentzian frame $\vartheta_\mu ^\al$ is to be treated  as the
dynamical one,
while the background  affine connection $\hat\Ga_{\mu\nu}^\la$   as  
nondynamical. Otherwise, under extremizing $S$, the
background   connection  should not be varied, $\de  \hat\Ga_{\mu\nu}^\la =0$.
 Under the requirement of  the
background-independence, the
action $S$ would preserve the  general diffeomorphism invariance/relativity. 
In the case of  the residual  dependence on  the background $
\hat\Ga_{\mu\nu}^\la$,  the action, though retaining the  general covariance, 
violates, partially or completely,  the general  invariance/relativity.

\subsubsection{Emergent metric}

Restricting himself  by  the pure gravity 
one can equivalently 
choose as an independent variable for gravity,  instead of the Lorentzian 
frame $\vartheta_{\mu}^\al$,  its   bilinear  Lorentz-invariant combination
\be
g_{\mu\nu}(x)=\vartheta_\mu^\al
\eta_{\al\beta}\vartheta_\nu^\beta= {e} _\mu^{\bmu} \vartheta_{\bmu}^\al
\eta_{\al\beta}   \vartheta_{\bnu}^\beta  {e} _\nu^{\bnu}=
{e} _\mu^{\bmu} g_{\bmu\bnu} {e}_\nu^{\bnu},
\ee
with $ |\mbox{\rm det} (\vartheta_\mu^\al) =\sqrt{|{\rm
det}(g_{\mu\nu})|}$.\footnote{By this
token, one can use the convectional relation ${e} ^{\bmu}_\mu=
g^{\bmu\bnu}
g_{\mu\nu} {e} _{\bnu}^\nu$, etc.} 
At that, the frame $\vartheta_\mu^\al$, of which  $g_{\mu\nu}$  is composed, 
though being the primary gravity field, manifests 
itself  explicitly  only in interactions with matter (omitted here).
The tensor field  $g_{\mu\nu}$ may be treated as an emergent metric.
It is the emergence of metric,
which converts a  background manifold  ${\cal M}_d$ with an affine connection
into the true space-time, the latter  becoming  in a sense  emergent,  as
well.\footnote{In
the same vein, one could formally consider  other  patterns of the affine
symmetry
breaking, $GL(d,R)\to SO(n,d-n)$, $n=0, 1, \dots, [d/2]$, resulting in
the same number of the affine Goldstone bosons and  the emergent 
metric with the space-time signature $(n,d-n) $, to be eventually 
selected~\cite{Pir1}. For $d=4$, cf., e.g.~\cite{Ng}.}

\section{Violated relativity and dark matter}

\subsection{Affine symmetry}

Building the proper  nonlinear model starts out   from  ${\cal R}_d$ in the
globally
affine
coordinates $\xi^m$.  To construct  the Lagrangian dependent on
the affine tensor  $g_{\bm\bn}$  and its derivatives  construct 
first the Christoffel-like affine tensor 
\be
\Ga_{\bm\bn}^{\bl} (\xi)
=\frac{1}{2}g^{\bl\bk}(\partial_{\bm} g_{\bn\bk}+ 
\partial_{\bn} g_{\bm\bk} - \partial_{\bk}  g_{\bm\bn}).
\ee
A derivative of $g_{\bm\bn}$  may uniquely be expressed through 
a combination of $\Ga_{\bm\bn}^{\bl}$ (and v.v.).
By means of the latter,  one can construct the Riemann  tensor
$R_{\bm\br\bn}^{\bl}$,
the Ricci tensor $R_{\bm\bn}=R_{\bm\bl\bn}^{\bl}$ and the Ricci
scalar $R=g^{\bm\bn} R_{\bm\bn}$.  Altogether, 
one can construct on ${\cal R}_d$ the generic affine-invariant 
second-order effective  Lagrangian for gravity
\be\label{secord}
L_g= \frac{1}{2}\ka^2_g \Big (L_0+ \sum_{i=1}^5   \ve _i  \Delta L_i \Big),
\ee 
with  the partial bi-linear in $\Ga^l_{mn}$  contributions as  follows:
\bea\label{L}
L_0=2\La   - R(g_{\bm\bn},  \Ga_{\bm\bn}^{\bl} ), &&   \Delta   L_1=
g^{\bm\bn
}\Ga_{\bm\bk}^{\bk}
\Ga_{\bn\bl}^{\bl},  \nn\\
 \Delta L_2= g_{\bm\bn}  g^{\bk\bl}g^{\br\bs} \Ga_{\bk\bl}^{\bm}  
\Ga_{\br\bs}^{\bn}, &&
 \Delta L_{3}=g^{\bm\bn} \Ga_{\bm\bn}^{\bk} \Ga_{\bk\bl}^{\bl},\nn\\
 \Delta  L_4=g_{\bm\bn}g^{\bk\bl} g^{\br\bs}    \Ga_{\bk\br}^{\bm}
\Ga_{\bl\bs}^{\bn}, &&
 \Delta  L_5= g^{\bm\bn} \Ga_{\bm\bk}^{\bl} \Ga_{\bn\bl}^{\bk} .
\eea
For completeness, there is included into $L_0$   a constant $\La$. 
The parameter $\ka_g=1/\sqrt{8\pi G}$ is the Planck mass, with $G$ being
the Newton's constant, and  $\ve_i$, $i=1,\dots,5$, are  the   
dimensionless free parameters. 
The presented  terms
exhaust all the bilinear in $\Ga^l_{mn}$ second-order  ones admitted  by the
affine
symmetry, with 
no  prior preference  among   them.\footnote{On the affine  symmetry reason,
one could add two linear 
terms, $g^{\bm\bn}\partial_{\bl} \Ga_{\bm\bn}^{\bl} $
and    $g^{\bm\bn}\partial_{\bm} \Ga_{\bn\bl}^l $, which however   may be
expressed  though  the rest of the terms  modulo surface contributions.}
Transforming the results to the curvilinear coordinates $x^\mu$ on ${\cal R}_d$
one would arrive at a generally covariant theory of gravity on a flat affine
background. However, this is just a limited  version  of a more general case
(see,
further on). 

\subsection{General covariance}

Let us then  map    the above results  from ${\cal R}_d$ onto  
${\cal M}_d$, first, in  the locally affine coordinates $ \xi^{\bmu}_X$
around $\Xi_X$ 
through the identical substitution
$\Ga^{\bl}_{\bm\bn}(\xi)\to\Ga^{\bl}_{mn}(\xi_X) $
and  then in the arbitrary observer's coordinates $x^\mu$ around~$X$ (using the
counterpart of   (\ref{trans}) for $\Ga_{mn}^l(\Xi_X)$ supplemented by 
(\ref{xbarx})). Altogether,    we  get the relation
\be
 \Ga_{\bmu\bnu}^{\bla} (\Xi_X)  ={e} _{\bmu}^\mu {e} _{\bnu}^\nu
{e} ^{\bla}_\la(X)\Big(  \Ga_{\mu\nu}^\la(X)  - \hat
\Ga_{\mu\nu}^\la(X) \Big),
\ee
where conventionally
\be\label{L}
\Ga_{\mu\nu}^\la(x) = \frac{1}{2} g^{\la\rho}(\partial_\mu g_{\nu\rho}+ 
\partial_\nu g_{\mu\rho}  -
\partial_\rho g_{\mu\nu})
\ee
(with $X\to x$) is the Christoffel connection corresponding to metric
$g_{\mu\nu}(x)$ .
Altogether, the most general  second-order  generally covariant effective
Lagrangian for the emergent metric gravity is given by
(\ref{secord}), with\footnote{The  terms  without  derivatives
of $g_{\mu\nu}$, as well as the extra factors given by  the powers of
$g=\det(g_{\mu\nu})$,  are  forbidden by  
the hidden affine symmetry.} 
\bea\label{Lg}
L_0=  2\La- R(g_{\mu\nu},  \Ga_{\mu\nu}^\la ), 
 &&
\Delta  L_1= g ^{\mu\nu} B_{\mu\ka}^\ka  B_{\nu\la}^\la   ,\nn\\
 \Delta L_2=g_{\mu\nu}  g^{\ka\la}g^{\rho\si}
B_{\ka\la}^\mu   B_{\rho\si}^\nu,&&
\Delta  L_{3}=g^{\mu\nu} B_{\mu\nu}^\ka B_{\ka\la}^\la     ,\nn\\
\Delta  L_4 = g_{\mu\nu}  g^{\ka\la}  g^{\rho\si} B_{\ka\rho}^\mu
B_{\la\si}^\nu , && 
\Delta L_5=  g^{\mu\nu}  B_{\mu\ka}^\la  B_{\nu\la}^\ka,
\eea
dependent on the generally covariant tensor\footnote{On the covariance
reason, such a dependence  was postulated in~\cite{Pir3}.}
\be\label{B}
B_{\mu\nu}^{\la}(x)\equiv\Ga_{\mu\nu}^\la - \hat\Ga_{\mu\nu}^\la .
\ee
The background-independent term $L_0$ corresponds to GR  with a
cosmological constant $\La$, while  the background-dependent  ones $\De L_i$,
  $i=1,\dots,5$, to the GR violation.\footnote{At that, under
the restriction  ab initio by GR, the flat affine background would
superficially suffice.}
The theory of gravity  given by  the GR-violating effective Lagrangian
(\ref{Lg}) and  (\ref{B})   may be referred to as Violated Relativity (VR). 
Most generally, it depends  on  the $d^2(d+1)/2$ nondynamical  functions 
besides the five constant Lagrangian parameters $\ve_i$.\footnote{Treating $\De
L_i$ as the small perturbations and  $L_0$ as  the leading term, one can extend
the latter by the higher-order generally covariant contributions.}

\subsection{Gravitational DM}

Varying the gravity action with respect  to $g_{\mu\nu}$, under fixed
$\hat\Ga_{\mu\nu}^\la$,  we get  
the vacuum gravity field equations in a generic form as follows: 
\be\label{FE}
R_{\mu\nu}-\frac{1}{2} R g_{\mu\nu}+ \La g_{\mu\nu}=\frac{1}{\ka_g^2}\sum_i^5
\ve_i\Delta T^{(i)}_{\mu\nu}(g_{ \rho\si},  B^\la_{\rho\si}) \equiv  
\frac{1}{\ka_g^2}\De T_{\mu\nu} .
\ee
Here $\Delta T^{(i )}_{\mu\nu}$ are  the generally covariant contributions
to the equations  due to  $\Delta L_i$: 
\be
\Delta T^{(i )\mu\nu}=\frac{2} {\sqrt{|g}|    }\frac{\de\Big (\sqrt{|g}| \De
L_i\Big)}{\de g_{\mu\nu}},
\ee
with $g=\det(g_{\mu\nu})$ and $\de/\de g_{\mu\nu}$ designating the total
variational derivative. The r.h.s.\ of (\ref{FE}) may formally be treated
as the covariantly conserved, $\na_\mu \De T^{\mu\nu}=0$,  canonical
energy-momentum tensor of the gravitational DM
due  to the GR violation ($\ve_i\neq0$). The affine  ``texture'' of
space-time with $\hat \Ga^\la_{\mu\nu}$, mimicking such  DM, 
signifies ultimately  the gravity and space-time as emergent.

\subsection{Limited GR  violation} 

Generally, the phenomenological study of   VR is 
rather cumbersome. To simplify it as much as possible,  consider the formal
limit $\hat\Ga_{\mu\nu}^\la = 0$
 corresponding to  the flat affine   background 
in the globally affine coordinates. 
The various observations being in reality   fulfilled in  the 
different coordinates, it is practically impossible for an
observer to guess/use the unknown  globally affine coordinates ab initio. In the
lack of this knowledge,
one should start  from suitable  observer's coordinates $x^\mu$, assuming  
some $\hat \Ga_{\mu\nu}^{\la} (x)$,  to
eventually  reveal, with  the help
 of observations,  the globally  affine coordinates
$\xi^{\bmu}$ (independent of any reference point $X$), so that  $\hat
\Ga_{\bm\bn}^{\bl} (\xi)\equiv 0$.
According to~(\ref{trans}),  one should have in this case  the restriction
\be\label{Ga}
\hat\Ga_{\mu\nu}^\la (x) 
=\frac{ \partial^2 
\xi^{\bla}}{\partial x^\mu\partial x^\nu} 
 \frac{ \partial x^\la}{\partial 
\xi^{\bla}}\bigg|_{\xi= \xi(x)}
 ={\hat e}^\la_{\bla} \partial_\mu{\hat e}^{\bla}_\nu .
\ee
Here  ${\hat e}^{m}_\mu=\partial
\xi^{m}/\partial x^\mu$ and ${\hat e}^\la_{\bla}= \partial x^\la /\partial
\xi^{l}|_{\xi= \xi(x)} $ are the frames relating  the distinguished globally
affine coordinates  and  the arbitrary observer's ones in terms of   the 
$d$
nondynamical generally covariant scalar fields $ \xi^{m}(x)$ 
(having the  inverse  $x^\mu(\xi)$). 
This is   the least set of free  functions to consistently account for  the 
GR violation, under preserving general
covariance. The topology  of the affine background, flat vs.\  curved,   is
thus not a pure theoretical question  but becomes, in principle,  liable to
observational verification.  Anyhow, the flat affine background 
determined  by  (\ref{Ga})
could  be treated as an  
approximation, in a space-time region,    to  a curved affine background 
determined by a regular $\hat\Ga^\la_{\mu\nu}$.\footnote{A  limiting generally
noncovariant  case corresponding here, in this paper,   to GR
violation with $\hat\Ga_{\mu\nu}^\la =  0$,  $\xi^m=\de^m_\mu
x^\mu$  and $\hat e_\mu^m=\de_\mu^m$   was elaborated, irrespective of
DM,   in~\cite{Donoghue}.} 

\subsection{Unimodular relativity} 
 
For  $\De L_1$  one has
\be
B_{\mu\la}^\la =\partial_\mu
\ln \sqrt{|g|} -\hat\ga_\mu,
\ee
where $\hat\ga_\mu\equiv \hat \Ga^\la_{\mu\la}$. 
Moreover, if  the affine background is flat,  one gets in view of (\ref{Ga}):
\be
\hat\ga_\mu =  \hat e^\la_l \partial_\mu \hat e_\la^l = \partial_\mu
\ln |\det( \hat e_\la^l   )|.
\ee
Thus under  this limitation one has
\be
\Delta L_1= g^{\mu\nu}\partial_\mu \vsi \partial_\nu \vsi,
\ee
where 
\be
 \vsi = \ln(\sqrt{ | g|}/\hat\mu ),
\ee 
with  
\be
 \hat\mu= |\det(\partial_\mu\xi^m)|.
\ee 
The field $\vsi$, determined by the ratio of the
two scalar densities of the same weight, behaves like a generally covariant 
scalar.  The nondynamical  field $\hat\mu$ being a unimodular scalar, the
theory with only $\Delta L_1$  (in addition to  $L_0$) may for uniformity be
referred to as 
Unimodular Relativity (UR),  with the
scalar graviton $\vsi$ serving  as the gravitational
DM~\cite{Pir1a}--\cite{Pir4}.
Now, the scalar graviton  is nothing but  a Goldstone boson
corresponding to  the hidden dilatation symmetry.\footnote{Such a
specific dilaton in disguise, representing  
a compression gravity mode in metric,  may be called the
``systolon''~\cite{Pir2}.} Besides, the so-called
``modulus''  $\hat\mu$ acquires  the  clear-cut physical meaning.\footnote{In
this particular case, it proves
that  the unknown  modulus $\hat \mu $ may   be hidden into
$\vsi$ taken as an independent   variable and, after finding the latter
together with  metric through  the field equations,  be
reconstructed as a consistency
condition. The proper solutions may be associated
with DM~\cite{Pir1a}--\cite{Pir4}. The
generally covariant formalism is
crucial to this point.}$ ^,$\,\footnote{A  theory of  
gravity  (the
so-called ``TDiff gravity``) in the  generally noncovariant form  corresponding
here, in this paper,  to UR  in  the gauge $\hat \mu=1$ 
was  elaborated, irrespective of DM,  in~\cite{Alv}.}
To qualitatively compile  with the astrophysical
data  on the galaxy anomalous  rotation curves due to  the dark
halos, there should fulfill $\ve_1^{1/2}\sim v_\infty\sim10^{-3}$,
with  $v_\infty$ being an asymptotic rotation velocity. 
To tame   phenomenologically the possible  unwanted  properties of the 
gravitational DM, a hierarchy of the GR violations
$|\ve_{2-5}|\ll |\ve_{1}|\ll1$ may be envisaged, with GR reduced first to UR 
and ultimately  to VR.

\section{Conclusion} 

 The nonlinear affine Goldstone model may
provide a prototype for the  emergent gravity,  with the GR violation and 
the gravitational DM serving as the signatures of  emergence.
Resulting in VR, given by   the Lagrangian
(\ref{Lg}) and (\ref{B})  (supplemented in the limited version by the
relation (\ref{Ga})),  the model  widely extends
the phenomenological  horizons beyond GR, with the possible reduction of GR 
first to UR and then
to VR. In an ultimate theoretical perspective, the model  may, hopefully, serve
as a guide towards a putative   underlying theory of  gravity and space-time.

\end{document}